\documentclass[aip,jap,amsmath,amssymb,
reprint,%
]{revtex4-2}

\usepackage{lmodern}
\usepackage[bookmarks,%
            colorlinks,%
            urlcolor=blue,%
            citecolor=blue,%
            linkcolor=blue,%
            hyperfigures,%
            pdfcreator=KN,%
            breaklinks=false]{hyperref}
\usepackage[english]{babel}
\usepackage{siunitx}
\usepackage{csquotes}
\usepackage{nth}
\usepackage{multirow}
\usepackage[version=4]{mhchem}
\usepackage[caption=false]{subfig}
\usepackage{graphicx}
\usepackage{orcidlink}
\DeclareGraphicsExtensions{.pdf}

\begin{document}

\title{Subcoercive-field dielectric response of \ce{0.5(Ba_{0.7}Ca_{0.3}TiO_{3})}--\ce{0.5(BaZr_{0.2}Ti_{0.8}O_{3})} thin film: peculiar third harmonic signature of phase transitions and residual ferroelectricity}
\author{Kevin Nadaud\orcidlink{0000-0002-2969-1453}}\altaffiliation{Author to whom correspondence should be addressed: \href{mailto:kevin.nadaud@univ-tours.fr}{kevin.nadaud@univ-tours.fr}\\ The following article has been submitted to Applied Physics Letters. After it is published, it will be found at \href{https://dx.doi.org/10.1063/5.0182718}{https://dx.doi.org/10.1063/5.0182718}}
\affiliation{%
GREMAN UMR 7347, Université de Tours, CNRS, INSA-CVL, 16 rue Pierre et Marie Curie, 37071 Tours, France%
}%
\author{Guillaume F. Nataf\orcidlink{0000-0001-9215-4717}}
\affiliation{%
GREMAN UMR 7347, Université de Tours, CNRS, INSA-CVL, 16 rue Pierre et Marie Curie, 37071 Tours, France%
}
\author{Nazir Jaber}
\affiliation{%
GREMAN UMR 7347, Université de Tours, CNRS, INSA-CVL, 16 rue Pierre et Marie Curie, 37071 Tours, France%
}
\author{Micka Bah\orcidlink{0000-0001-6636-2854}}
\affiliation{%
GREMAN UMR 7347, Université de Tours, CNRS, INSA-CVL, 16 rue Pierre et Marie Curie, 37071 Tours, France%
}
\author{Béatrice Negulescu\orcidlink{0000-0001-7218-9859}}
\affiliation{%
GREMAN UMR 7347, Université de Tours, CNRS, INSA-CVL, 16 rue Pierre et Marie Curie, 37071 Tours, France%
}
\author{Pascal Andreazza\orcidlink{0000-0001-8042-5583}}
\affiliation{%
ICMN, CNRS, Université d'Orléans, 1b rue de la Férollerie, CS 40059, 45071 Orléans Cedex 02, France%
}
\author{Pierre Birnal}
\affiliation{%
ICMN, CNRS, Université d'Orléans, 1b rue de la Férollerie, CS 40059, 45071 Orléans Cedex 02, France%
}
\author{Jérôme Wolfman}
\affiliation{%
GREMAN UMR 7347, Université de Tours, CNRS, INSA-CVL, 16 rue Pierre et Marie Curie, 37071 Tours, France%
}


\begin{abstract}
    Sub-coercive field non-linearities in \ce{0.5(Ba_{0.7}Ca_{0.3}TiO_{3})-0.5(BaZr_{0.2}Ti_{0.8}O_{3})} (BCTZ 50/50) thin film elaborated using pulsed laser deposition are studied using permittivity and phase angle of the third harmonic measurements as function of the AC measuring field $E_{\mathit{AC}}$ and temperature.
    The global phase transition temperature $T_{\mathit{max}}$ for which the permittivity is maximum, decreases from \qty{330}{K} to \qty{260}{K} when $E_{\mathit{AC}}$ increases.
    Rayleigh analysis of the AC field dependence of the relative permittivity shows a regular decrease of the domain wall motion contributions as temperature increases up to $T_{\mathit{max}}$ and an even more pronounced decrease above $T_{\mathit{max}}$.
    This measurement reveals that the ferroelectric behavior subsists \qty{70}{\K} above the global phase transition.
    The phase angle of the third harmonic at temperatures below \qty{275}{\K}, is characteristic of a conventional ferroelectric and from \qty{275}{\K} to $T_{\mathit{max}}=\qty{330}{\K}$ of a relaxor.
    Above $T_{\mathit{max}}$, the thin film exhibits a peculiar phase angle of the third harmonic, which consists of $\qty{-180}{\degree}\rightarrow \qty{-225}{\degree}\rightarrow \qty{+45}{\degree} \rightarrow \qty{0}{\degree}$ instead of the $\qty{-180}{\degree}\rightarrow \qty{-90}{\degree} \rightarrow \qty{0}{\degree}$ found for relaxor.
    This peculiar behavior is observed only on heating, and is tentatively attributed to changes in the correlations between polar nanoregions.
\end{abstract}
\maketitle

Relaxor ferroelectrics are promising materials for energy storage devices and actuators thanks to their high dielectric permittivity and high piezoelectric coefficients \cite{JayakrishnanPMS2023,VeerapandiyanM2020}.
In this context, \ce{BaTiO3}-based materials such as the \ce{Ba_{1-x}Ca_{x}Ti_{1-y}Zr_{y}O3} (BCTZ) solid solution represent an interesting alternative to lead-based materials \cite{SimonJACLCOM2018,DaumontJAP2016,PuliJPD2019,YanJMC2020}.
Their properties are strongly linked to the dynamics of domain walls, in their ferroelectric phases, that can enhance the electromechanical response but also significantly increase losses \cite{ZhengAFM2023,CarpenterJPCM2015,DamjanovicJACS2005}.
In addition, they exhibit polar nanoregions (PNRs), clusters in which the polarization is randomly aligned in absence of external electric field\cite{CollaJAP1998,CowleyAP2011,BokovJMS2006,OtonicarAFM2020}, that contribute significantly to the macroscopic dielectric and piezoelectric responses \cite{Li2016,LiAFM2018,LiAFM2017}.

Those PNRs can subsist several kelvins above the Curie temperature (in the centrosymmetric phase) \cite{GartenJACS2016,GartenJAP2014,SaljePRB2013} and are visible in the non-linear response of the materials, similar to the signature of domain walls in conventional ferroelectrics \cite{OtonicarJACS2022,OtonicarAFM2020,HashemizadehAPL2017}.
On cooling from the paraelectric phase, relaxor ferroelectrics can be described through three key temperatures\cite{LiAFM2018,MaitiF2011}: (1) the Burns temperature $T_{B}$ where the population of PNRs begins to be significant \cite{BurnsSSC1973}, (2) the freezing temperature $T_{f}$ where the PNRs have grown such that they become static \cite{PircPRB2007}, (3), the depolarization temperature $T_{d}$ where long-range ferroelectric domains can be achieved by electric-field poling \cite{DkhilJAP2001,RaevskiPRB2005}, induced by the percolation of PNRs\cite{ProsandeevPRB2013}.

In this article, we study the phase transitions in a thin film of BCTZ 50/50 using the relative permittivity at the first harmonic (using hyperbolic law) and the evolution of phase angle of the third harmonic as a function of the AC field for different temperatures.
Using hyperbolic analysis, we show that residual ferroelectricity persists up to \qty{70}{\K} above the maximum in permittivity indicating the transition to the cubic phase.
Above this temperature, a faster decay of the reversible and irreversible domain wall motion contributions is visible.
The crossing of this temperature is also well visible on the phase angle of the third harmonic, which exhibits a totally different evolution with increasing AC field below and above it.

At sub-switching AC fields, for a homogeneous distribution of pinning centers, the relative permittivity can be described using the Rayleigh law:\cite{taylorjap1997,SchenkPRA2018}
\begin{equation}
    \varepsilon_{r} = \varepsilon_{\mathit{r-l}} + \alpha_{r}E_{\mathit{AC}}
    \label{rayleigh,}
\end{equation}
Where $E_{\mathit{AC}}$ corresponds to the magnitude of the applied measuring electric field, $\varepsilon_{\mathit{r-l}}$ is the lattice contribution to the permittivity, $\alpha_{r}$ is the irreversible contribution from the motion of domain walls (domain wall pinning/unpinning), polar cluster boundaries, or phase boundaries and corresponds to the slope of the permittivity v.s. electric field curve.
In such conditions of homogeneous distribution of pinning centers, the polarization versus electric field loop can be described using the following expression:\cite{taylorjap1997,OtonicarJACS2022}
\begin{equation}
    \label{eq:polarization}
    P = \varepsilon_{0}\left(\varepsilon_{\mathit{r-l}}+\alpha_{r}E_{0}\right)E \pm \frac{\varepsilon_{0}\alpha_{r}}{2}\left(E_{0}^{2}-E^{2}\right) + \dots
\end{equation}
The sign $+$ stands for the decreasing and the sign $-$ for the increasing part of the AC field.
The second term reflects the hysteretic contribution of domain walls to the polarization.
This non-linear expression of the polarization gives the following Fourier series decomposition when the applied electric field is $E(t) = E_{0}\sin\left(\omega t\right)$:\cite{RiemerJAP2021}
\begin{align}
    P(t, E_{0}) = \varepsilon_{0}\left(\varepsilon_{\mathit{r-l}}+\alpha_{r}E_{0}\right)E_{0}\sin\left(\omega t\right)\nonumber \\ + \sum_{1,3,5,\dots} \frac{4\varepsilon_{0}\alpha_{r}E_{0}^{2}\sin\left(\frac{\pi n}{2}\right)}{\pi n(n^{2}-4)}\cos\left(n\omega t\right)
    \label{eq:fourier rayleigh}
\end{align}
The irreversible domain wall motion contribution is thus out-of-phase with the measuring electric field in the case of an ideal material.
In order to describe a real material, equation \eqref{eq:polarization} can contain additional terms, reflecting the degree of randomness of the energy profile,\cite{HashemizadehAPL2017} and in that case, harmonics may not be purely out-of-phase.
For this reason, the non-linear response of a relaxor or a ferroelectric material can be investigated by extracting the phase of the third-harmonic contribution to the polarization and its evolution with the measuring field amplitude, which gives information on the hysteretic or non-hysteretic character of domain wall motions\cite{OtonicarJACS2022,OtonicarAFM2020,MorozovJAP2008,HashemizadehAPL2017,NadaudJAP2023}.

In a real material, the distribution of pinning centers is not homogeneous and for low AC fields, the relative permittivity is almost constant, corresponding to reversible domain wall contributions, also called domain wall vibration\cite{SchenkPRA2018,BassiriGharbJE2007}.
A generalized expression can then be used to describe the permittivity evolution, called the hyperbolic law:\cite{borderonapl2011,NadaudAPL2021,BaiCI2017}
\begin{equation}
    \varepsilon_{r} = \varepsilon_{\mathit{r-l}} + \sqrt{\varepsilon_{\mathit{r-rev}}^2 + (\alpha_{r}E_{\mathit{AC}})^2}
    \label{hyperbolic}
\end{equation}
with $\varepsilon_{\mathit{r-rev}}$ the reversible domain wall motion contribution, proportional to the domain wall density\cite{boserjap1987,NadaudAPL2022,BorderonSR2017}.
$\varepsilon_{\mathit{r-l}}$, $\varepsilon_{\mathit{r-rev}}$ and $\alpha_{r}$ can be obtained by measuring the relative permittivity as a function of the driving field. 
Their evolution with driving frequency, temperature, DC bias field or previous states, allows understanding dielectric relaxations \cite{renouduffc2011}, residual ferroelectricity\cite{GartenJAP2014}, contributions to tunability\cite{NadaudJAP2016,GharbJAP2005} and annealing/cycling effects\cite{NadaudAPL2022,NadaudJAP2023}.

Irreversible domain wall motions make the relative permittivity AC field dependent, which is critical in many application\cite{GhoshAFM2013} especially because it can subsist in the microwave frequency range\cite{NadaudAPL2018}.

Polycristalline \qty{380}{nm} thick BCTZ film has been grown on \ce{Pt/TiO2 /MgO} substrate by pulsed laser deposition. 
Details on growth conditions can be found in the supplementary information.  
Top circular Au/Ti electrodes (\qty{150}{\um} radius) were deposited through a shadow mask. 
The metal-insulator-metal topology has been chosen for the simple extraction of the dielectric properties using the parallel plate capacitor formula.

X-ray diffraction (see diffraction pattern in Fig. S1 of supplementary information) shows that the BCTZ film is single-phased and polyoriented. 
Peak positions could be indexed according to orthorhombic BCTZ 50/50 (pdf \#04-022-8189). 
The film composition has been characterized by Rutherford Back Scattering (RBS, see Figure S2a,b) and was found to be close to the nominal target composition (see table S1) .

The single frequency Vertical Piezoresponse Force Microscopy (VPFM) is used to map the electrical polarization at nanoscale in the BCTZ film. 
A SCM-PTSI tip with a spring constant of \qty{2.8}{\N\per\m} and a radius of curvature of \qty{15}{nm} is mounted on the probe holder. 
An AC amplitude of \qty{0.5}{V} at a drive frequency of \qty{327}{kHz} (i.e. the contact resonance frequency) is applied to the BCTZ sample while the tip is grounded. 
The scan size and the scan rate are set to \qty{500}{nm} and \qty{0.2}{Hz}, respectively.

The dielectric characterizations presented in this article have been acquired using a lock-in amplifier (MFLI with MD option, Zurich Instrument) connected to a temperature-controlled probe station (Summit 12000, Cascade Microtech).
The AC measuring signal has been generated using the embedded lock-in generator.
Its amplitude has been swept from \qty{10}{mV_{rms}} to \qty{1}{V_{rms}} at a frequency of \qty{10}{kHz}.
The applied voltage and current through the capacitor are measured simultaneously by the lock-in.
The first harmonic of the voltage, in addition to first, second and third harmonics of the current, are demodulated simultaneously.
$|V|\exp\left(j\theta_{V}\right)$ is the phasor representing the applied voltage and $|I_{k}|\exp\left(j\theta_{I_{k}}\right)$ the phasor representing the $k$-th harmonic of the current (with $j$ the imaginary unit).
The first harmonic of the current and the applied voltage are used to compute the complex impedance:\cite{NadaudJALCOM2022,NadaudAPL2021,NadaudJAP2023}
\begin{equation}
    Z = \frac{|V|}{|I_{1}|}\exp\left(j\left(\theta_{V}-\theta_{I_{1}}\right)\right)
\end{equation}
The complex capacitance $C^{*}$ can hence be derived from the complex impedance:
\begin{equation}
  C^{*} = \frac{1}{j\omega Z} = \frac{|I_{1}|}{\omega|V|}\exp\left(j\left(\theta_{I_{1}}-\theta_{V} - \frac{\pi}{2}\right)\right)
\end{equation}
$\omega$ the angular frequency of the measuring voltage.
The material relative permittivity $\varepsilon_{r}^{*}$: 
\begin{equation}
  \varepsilon_{r}^{*} = \frac{t}{S\varepsilon_{0}}C^{*},
\end{equation}
with $t$ the thickness of the film, $S$ the surface of the electrodes and $\varepsilon_{0}$ the vacuum permittivity.
In the present case, the electrodes are sufficiently thick in order to limit the effect of the series resistance on the measured impedance.

In addition to the measurement as a function of the AC amplitude, the relative permittivity has been measured from \qty{31}{Hz} to \qty{31}{kHz} with an AC amplitude of \qty{40}{mV_{rms}}.

The $k$-th harmonic phase angle extraction process is similar and its value can be obtained using the following expression:\cite{NadaudJAP2023}
\begin{equation}
    \delta_{k} = \theta_{I_{k}}-k\theta_{V} - \frac{\pi}{2}
\end{equation}

\begin{figure*}
    \centering
    \subfloat{%
    \includegraphics[width=0.4\textwidth]{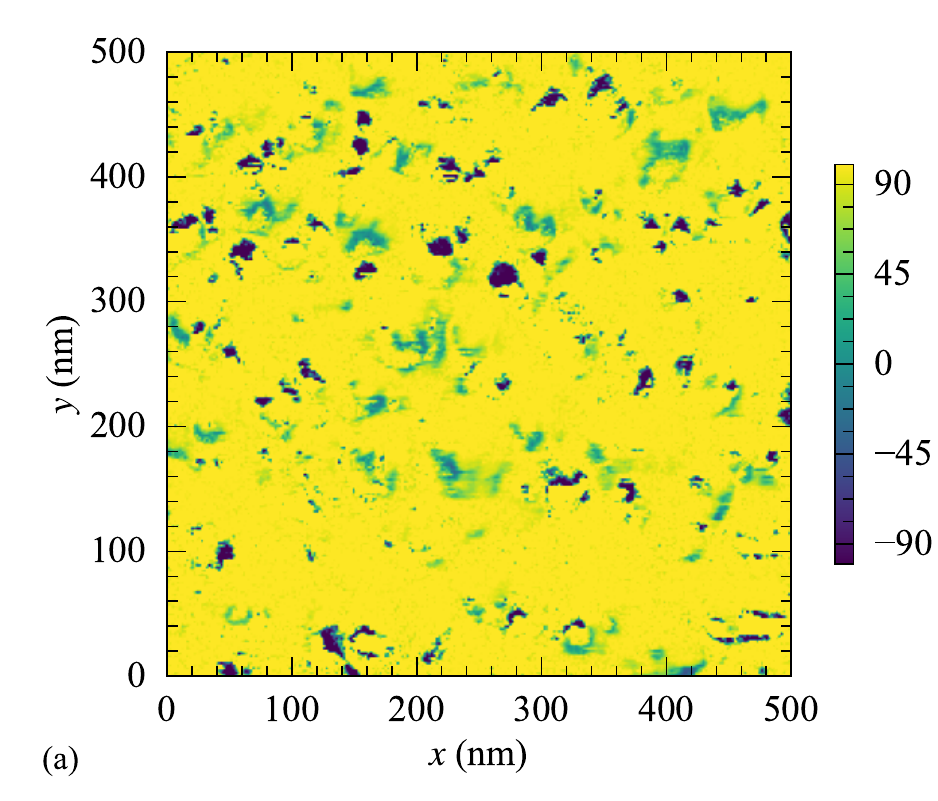}\label{subfig:PFM phase}}%
    \subfloat{%
    \includegraphics[width=0.4\textwidth]{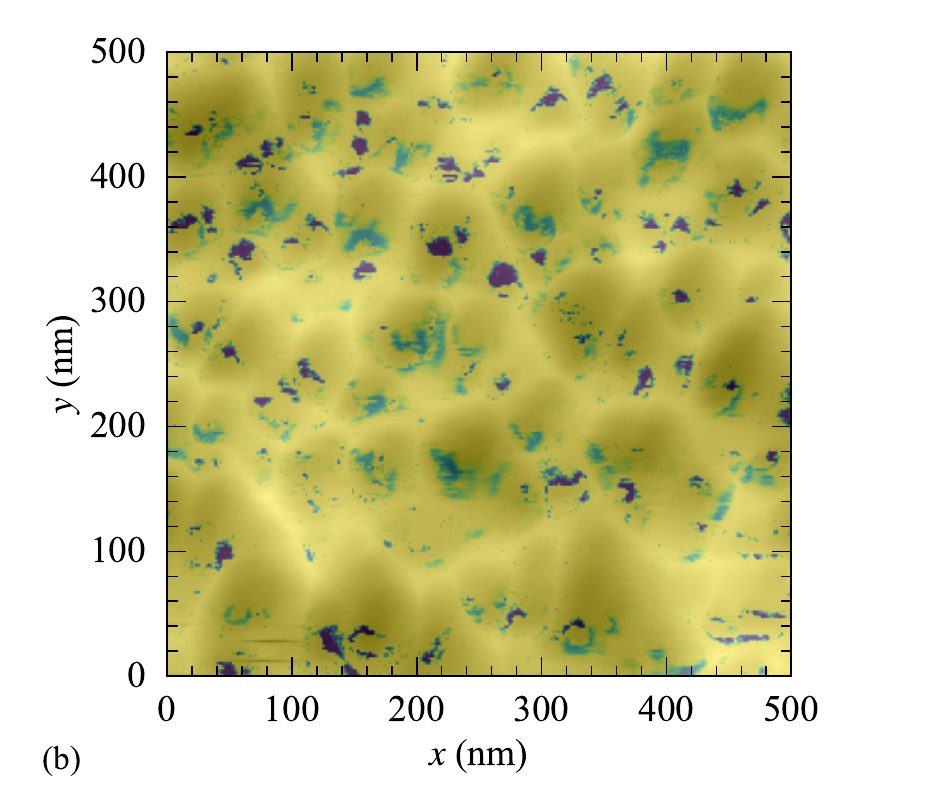}\label{subfig:PFM 3D phase}}%

    \subfloat{%
    \label{subfig:PFM cycle}\includegraphics[width=0.48\textwidth]{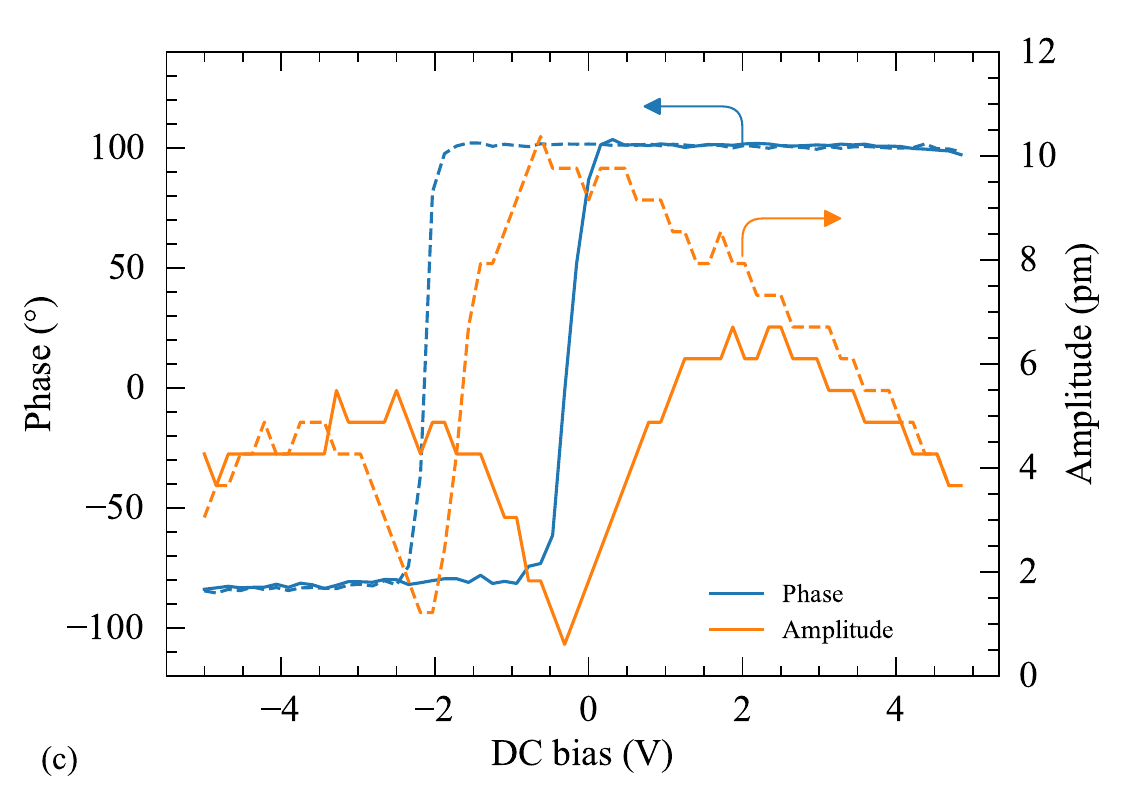}}
    \caption{%
    Phase of the Vertical PFM (VPFM) (a). 
    Superimposition of the phase signal and the topography (b). 
    PFM phase and amplitude as a function of an applied DC voltage $V_{DC}$ (c).
    }
    \label{fig:PFM}
\end{figure*} 

Fig.~\ref{subfig:PFM phase} displays the PFM phase, associated with the domain orientation (up or down). 
It highlights that the majority of the grains in the thin film form a single domain in the out-of-plane projection of the electrical polarization. 
Nevertheless, few grains present a polydomain configuration, as the superimposition of the topography one and that of the PFM phase shows (see dark and bright colors in Fig.~\ref{subfig:PFM 3D phase} ). 
Fig.~\ref{subfig:PFM cycle} displays the PFM phase and amplitude signals as a function of DC bias. 
These piezoresponse data are a signature of ferroelectric behavior. 
A phase difference of 182° between the up and down domains is observed at the coercive voltage ($|V_{c}| \simeq \qty{1}{V}$).
At the same time, the PFM amplitude drops as the polarization switching under the tip gradually takes place.
Topography measurement using AFM is given in supplementary material.

\begin{figure*}
    \centering
    {\includegraphics[width=0.48\textwidth]{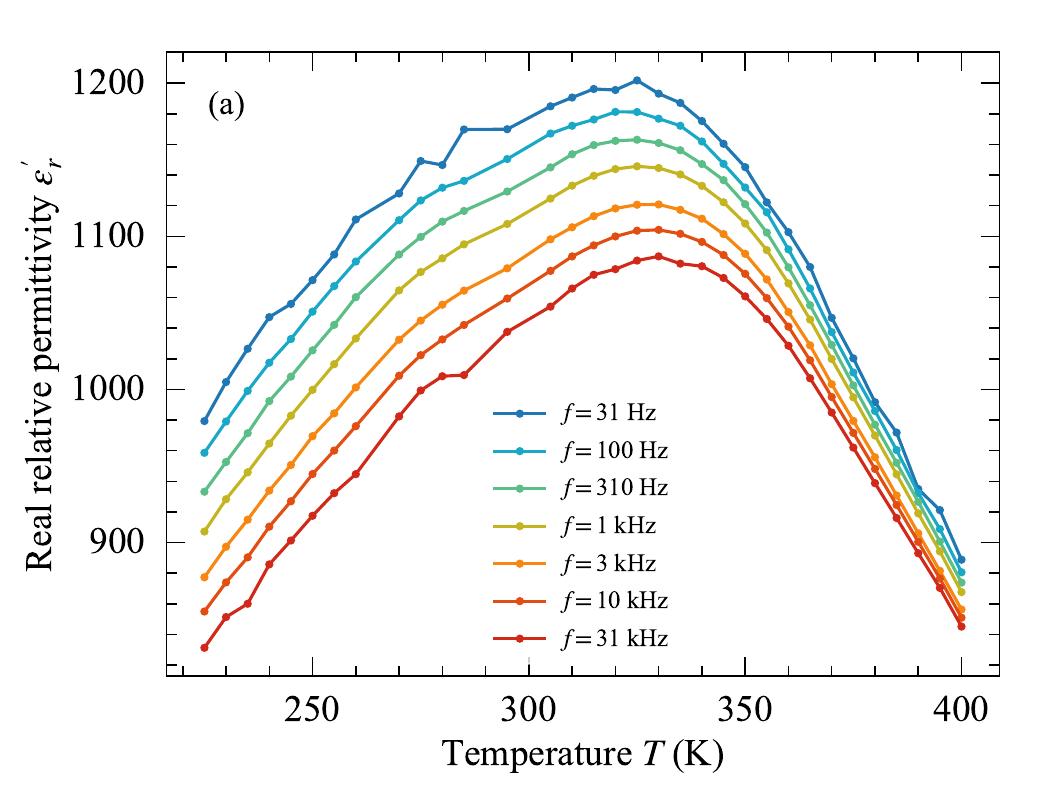}%
    \subfloat{\label{subfig:realPerm FREQ TEMP}}}%
    {\includegraphics[width=0.48\textwidth]{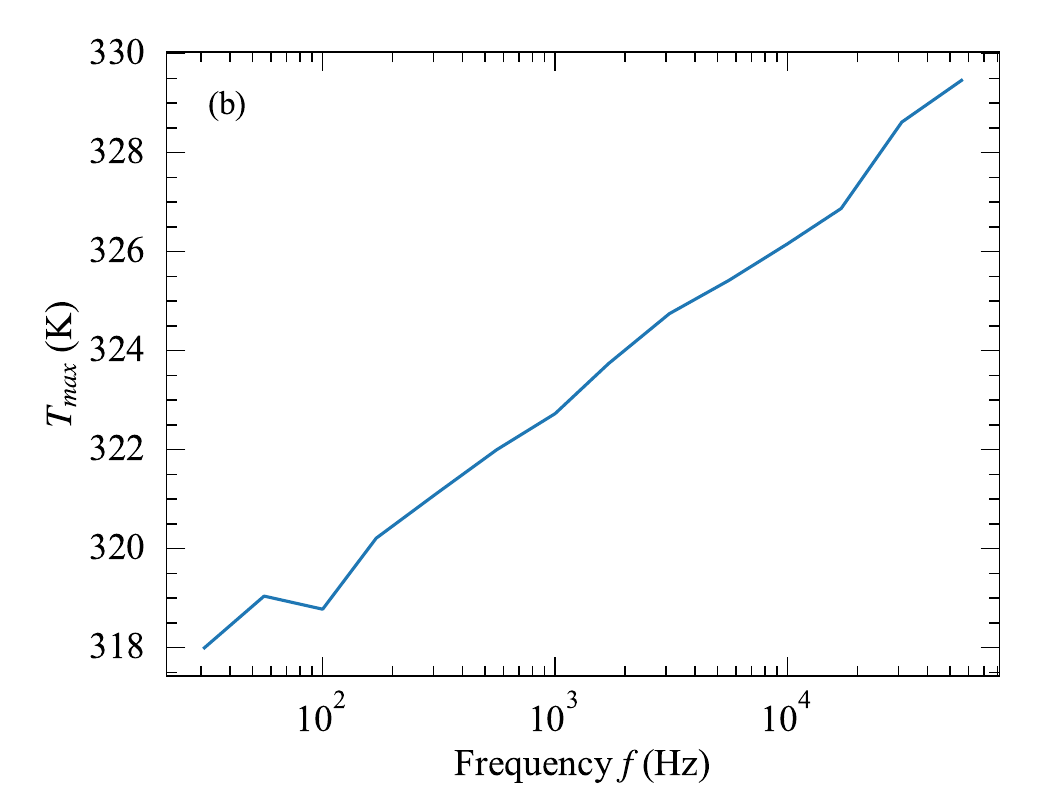}
    \subfloat{\label{subfig:realPermTmax FREQ}}}%
    {\includegraphics[width=0.48\textwidth]{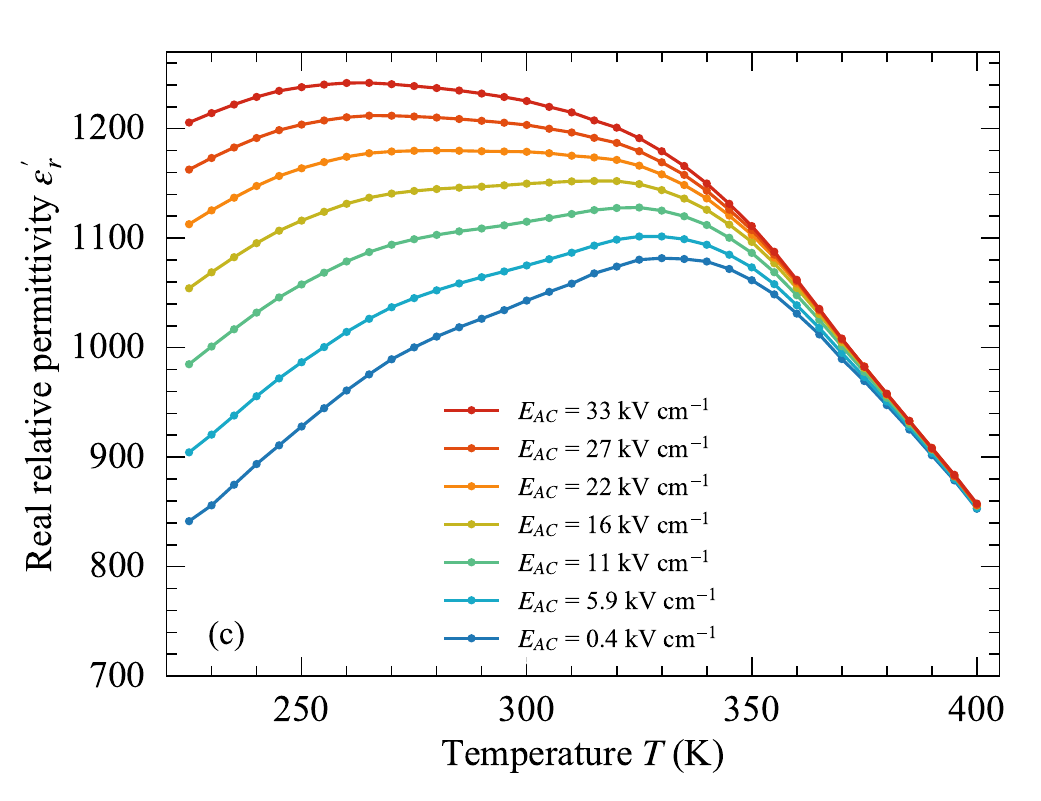}%
    \subfloat{\label{subfig:realPerm OLEV TEMP}}}%
    {\includegraphics[width=0.48\textwidth]{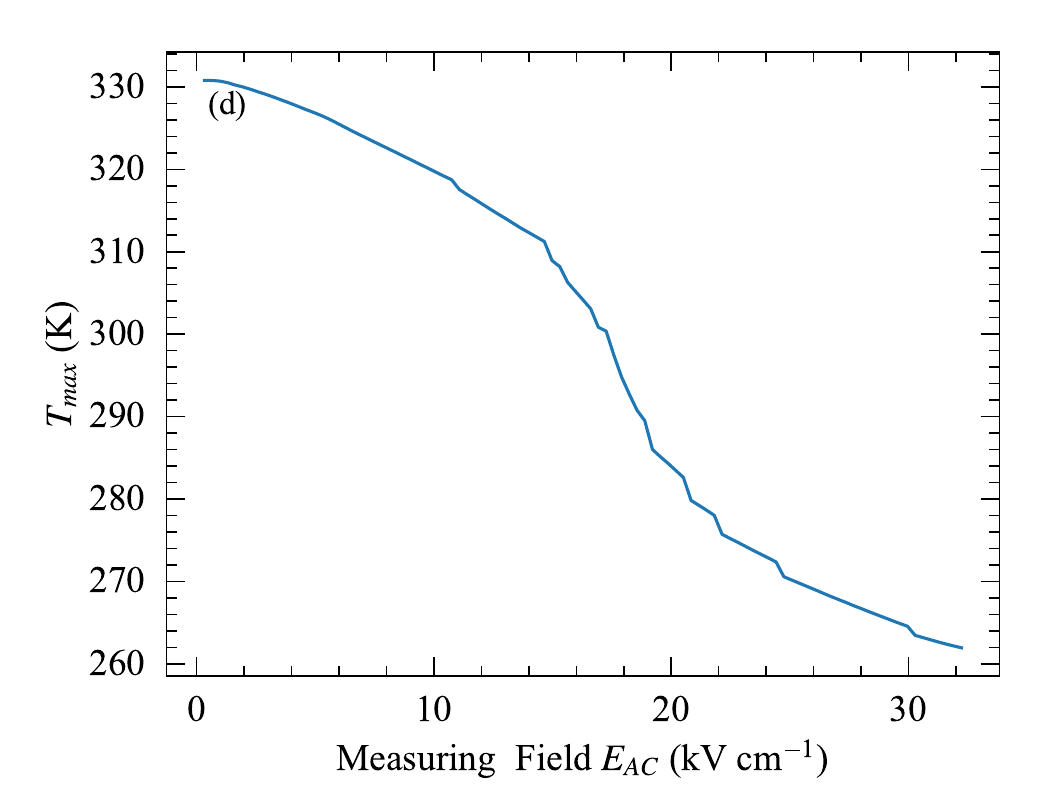}
    \subfloat{\label{subfig:realPermTmax OLEV}}}%
    \caption{%
    Relative permittivity as a function of the temperature for different measurement frequency for $E_{\mathit{AC}} = \qty{1.2}{\kV\per\cm}$ (a). %
    $T_{\mathit{max}}$ for which the relative permittivity is maximum as a function of the frequency for $E_{\mathit{AC}} = \qty{1.2}{\kV\per\cm}$ (b). %
    Relative permittivity as a function of the temperature for different values of the AC measuring field for $f=\qty{10}{kHz}$ (c). %
    $T_{\mathit{max}}$ for which the relative permittivity is maximum as a function of the AC measuring field for $f=\qty{10}{kHz}$ (d).%
    }
    \label{fig:realPerm OLEV TEMP}
\end{figure*}

Fig.~\ref{subfig:realPerm FREQ TEMP} shows the relative permittivity as a function of temperature for different frequencies.
$T_{\mathit{max}}$ have been determined from a least square fitting of $\varepsilon_{r}'$ with a parabola\cite{MaAPL2013}.
The decrease of the permittivity when the frequency increases is stronger for low temperatures (below \qty{300}{\K}) than for high temperatures (above \qty{340}{\K}).
This results in a shift of the maximum permittivity temperature $T_{\mathit{max}}$ with frequency (Fig.\ref{subfig:realPermTmax FREQ}), typical of relaxor-ferroelectrics behavior\cite{LiAFM2018,BokovJMS2006}.
The relaxor behavior is confirmed by the $P(E)$ loop and the analysis of the phase transition diffuseness, using modified Curie-Weiss analysis, both in supplementary material (Fig. S4 and S5).

Fig.~\ref{subfig:realPerm OLEV TEMP} shows the relative permittivity as a function of temperature for different values of the AC measuring field.
For an AC field of \qty{0.4}{\kV\per\cm}, $T_{\mathit{max}}$ is \qty{330}{K}, i.e. slightly lower than that reported for bulk BCTZ ceramics in the literature (\qty{363}{K}\cite{DamjanovicAPL2012}, \qty{365}{K}\cite{FengCI2019}, \qty{373}{K}\cite{GaoPSSA2020}) but very similar to the one obtained for thin films: \qty{323}{K}\cite{LiuCI2023,XuJACS2015}.
An inflection point is also visible, at \qty{270}{K}, and may correspond to a structural transition. 
It is close to the  tetragonal-orthorhombic and orthorhombic-rhombohedral transition temperatures observed in bulk ceramics \cite{BjrnetunHaugenJAP2013} and matches with anomalies observed in pyroelectric current \cite{BenabdallahJAP2011} and dielectric and elastic properties \cite{DamjanovicAPL2012}. 
Bulk \ce{0.5(Ba_{0.7}Ca_{0.3}TiO_{3})-0.5(BaZr_{0.2}Ti_{0.8}O_{3})} (BCTZ 50/50) is cubic (Pm-3m) above \qty{360}{\K}.
On cooling, it enters a tetragonal phase (P4mm), that coexist with a rhombohedral phase (R3m) between \qty{310}{\K} and \qty{210}{\K}.
Below \qty{210}{\K}, it is rhombohedral \cite{BjrnetunHaugenJAP2013}. 

When the AC measuring field increases, the relative permittivity increases due to irreversible contributions from the motion of domain walls, polar cluster boundaries, or phase boundaries \cite{CaiPRB2016,CollaJAP1999,ShettyAFM2019}.
One can note that the lower the temperature, the higher the increase of the relative permittivity with AC field.
The consequence of this increase is a shift towards lower temperatures of the temperature where the relative permittivity is maximum ($T_{\mathit{max}}$) when the AC field increases (Fig.~\ref{subfig:realPermTmax OLEV}).
Thus, in the following, the estimation of the maximum of the permittivity $T_{\mathit{max}}$ indicative of the transition to the cubic phase, is conducted at low AC field, to reduce the influence of domain wall motion.

The measurement has also been performed for decreasing temperature ($\qty{400}{K}\rightarrow \qty{225}{K}$) and presented in supplementary material (Fig.~S5).
The permittivity exhibits higher values and a maximum at a slightly lower temperature (\qty{320}{K}) compared to increasing temperatures (\qty{330}{K}).

\begin{figure*}
    \centering
    {\includegraphics[width=0.4\textwidth]{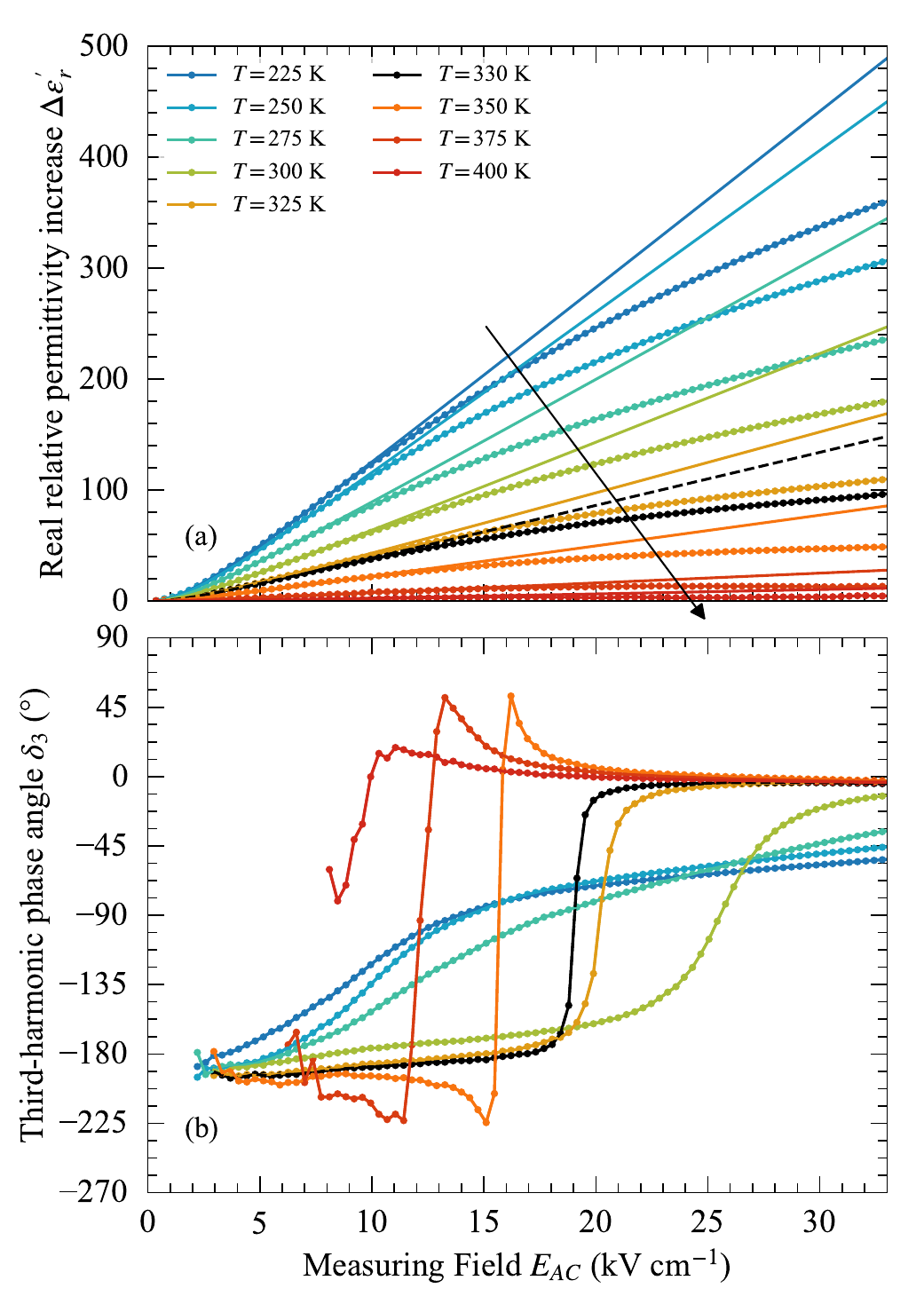}
    \subfloat{\label{subfig:realPerm TEMP OLEV inc}}%
    \subfloat{\label{subfig:delta TEMP OLEV inc}}}%
    {\includegraphics[width=0.44\textwidth]{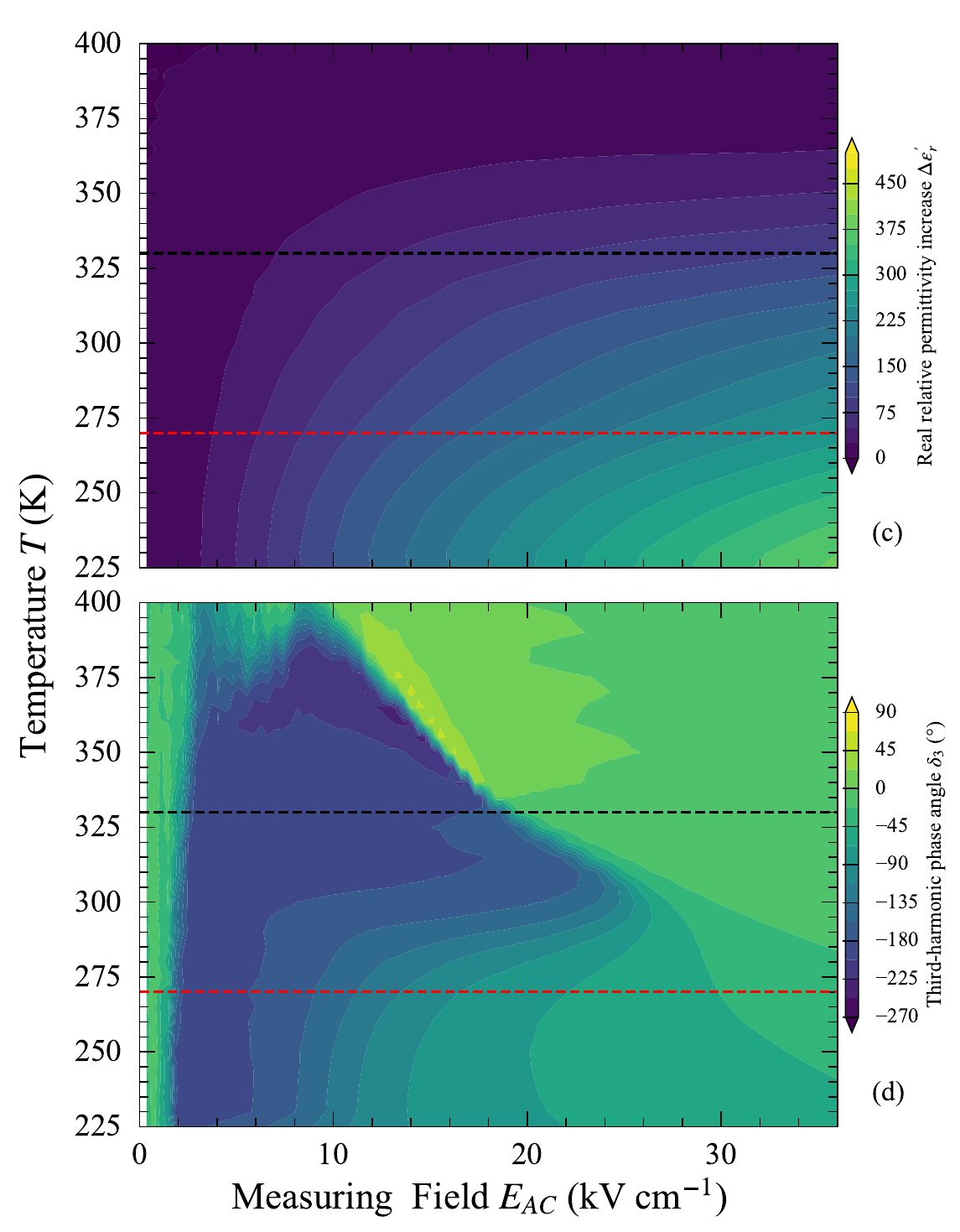}
    \subfloat{\label{subfig:realPerm TEMP OLEV map inc}}%
    \subfloat{\label{subfig:delta TEMP OLEV map inc}}}%
    {\includegraphics[width=0.4\textwidth]{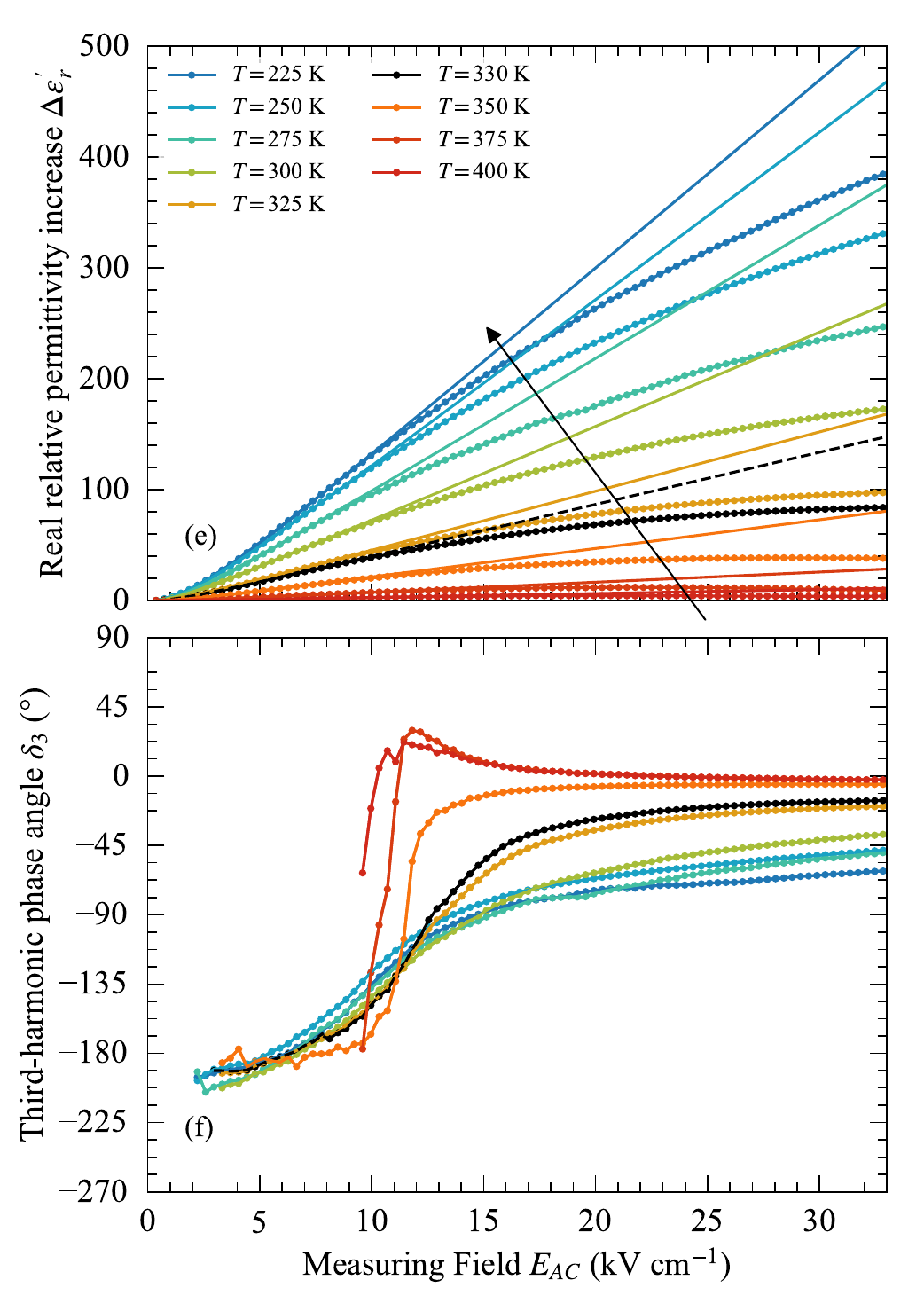}
    \subfloat{\label{subfig:realPerm TEMP OLEV dec}}%
    \subfloat{\label{subfig:delta TEMP OLEV dec}}}%
    {\includegraphics[width=0.44\textwidth]{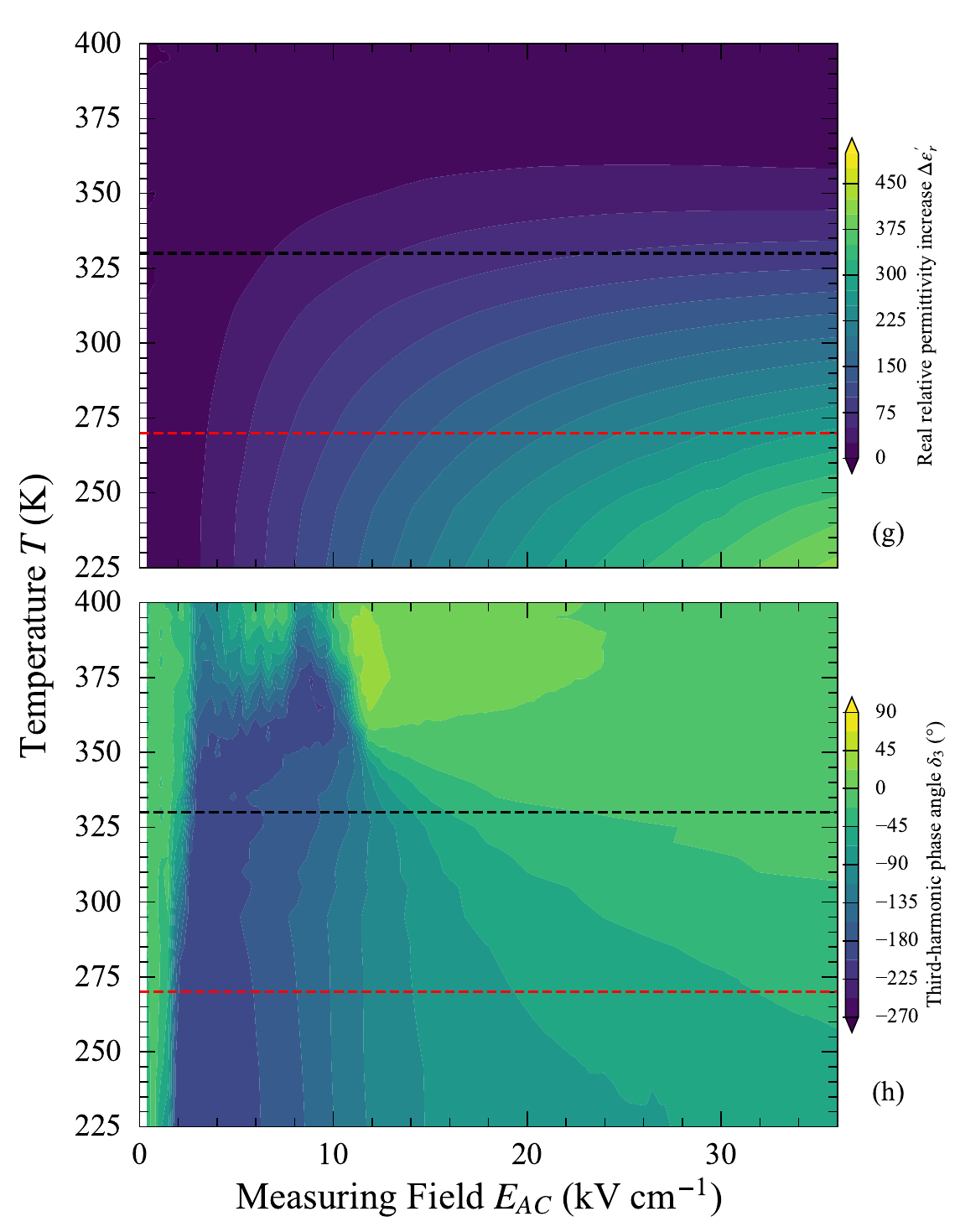}
    \subfloat{\label{subfig:realPerm TEMP OLEV map dec}}%
    \subfloat{\label{subfig:delta TEMP OLEV map dec}}}%
    \caption{%
    Relative permittivity variation (a,c) and phase angle of the third harmonic (b,d) as a function of the AC measuring field for different values of the temperature, on heating and (e,f,g,h) on cooling.
    For relative permittivity presented in (a), dots correspond to the experimental data and the solid lines to the hyperbolic fits.
    The black dashed line corresponds to $T_{\mathit{max}}$ (at $E_{\mathit{AC}}=\qty{0.4}{\kV\per\cm}$).
    The red dashed line in (c,d) corresponds to the inflection point at $T = \qty{270}{K}$.
    Data presented in (a,b,e,f) corresponds to slices every \qty{25}{K} of the data presented in (c,d,g,h). 
    }
    \label{fig:realPerm delta TEMP OLEV}
\end{figure*}

Fig.~\ref{fig:realPerm delta TEMP OLEV} shows the relative permittivity variation $\Delta \varepsilon_{r}' = \varepsilon_{r}' - \varepsilon_{r}'(0)$ and the phase angle of the third harmonic, as a function of the AC measuring field amplitude and for different temperatures, on heating.
$\Delta \varepsilon_{r}'$ increases when the AC field increases, due to the irreversible domain wall motion contribution. 
This increase is stronger at low temperatures (Fig.\ref{subfig:realPerm TEMP OLEV inc},\ref{subfig:realPerm TEMP OLEV map inc}).
The increase in permittivity vs AC field persists above $T_{\mathit{max}}$ (black dashed line), corresponding to residual ferroelectricity, similar to what has been observed in \ce{(Ba,Sr)TiO3}\cite{GartenJAP2014,GartenJACS2016,GartenAPL2014}.
Similar trends are observed when the measurement is performed on cooling (Fig.\ref{subfig:realPerm TEMP OLEV dec},\ref{subfig:realPerm TEMP OLEV map dec}).

Fig.~\ref{subfig:delta TEMP OLEV inc} shows the phase angle of the third harmonic as a function of the AC measuring field, for different temperatures, on heating. 
At low temperatures, up to \qty{275}{K}, the evolution of the phase-angle is typical of soft or weakly hard ferroelectrics\cite{OtonicarJACS2022}: for low AC fields, $\delta_{3}\simeq\qty{-180}{\degree}$, then when the AC field increases, $\delta_{3}$ increases and tends to a limit value around \qty{-45}{\degree}. 
This transition indicates a flat and random pinning potential for domain walls that can easily move both in reversible and irreversible ways. 
At low fields, a value of $\delta_{3}=\qty{-180}{\degree}$ corresponds to a non-hysteretic contribution of the domain wall motion to the permittivity\cite{RiemerJAP2021,OtonicarAFM2020}. 
This value is found for many ferroelectrics at low AC fields where the reversible domain motion contribution (vibration) dominates\cite{NadaudJAP2023,SchenkPRA2018,BassiriGharbJE2007}. 
At higher fields, the phase angle of the third harmonic evolves toward \qty{-90}{\degree} and corresponds to a purely hysteretic domain wall motion contribution to the permittivity, which results in a linear increase of the permittivity. 
It is the theoretical Rayleigh behavior. At even higher fields, a saturating-like response is observed with $\delta_{3}$ around \qty{-45}{\degree}. 
The evolution of the permittivity for these temperatures (Fig.~\ref{subfig:realPerm TEMP OLEV inc}) also corresponds to what is obtained for soft or weakly hard ferroelectrics: a very limited range for which the permittivity linearly increases (from \qty{3}{\kV\per\cm} to \qty{8}{\kV\per\cm}) then, the permittivity follows a sublinear increase.
When the temperature increases, the phase evolution with the AC field is similar up to \qty{275}{K}, the main difference resides in the phase-angle value at large fields (\qty{33}{\kV\per\cm}) which progressively increases. 

At \qty{300}{K}, the phase-angle increases from \qty{-180}{\degree} to \qty{-135}{\degree} and then increases quickly to \qty{0}{\degree}.
This type of phase-angle response is close to the one obtained for relaxors\cite{OtonicarJACS2022,OtonicarAFM2020,HashemizadehAPL2017,RiemerJAP2021}. 
A value of $\delta_{3}=\qty{0}{\degree}$  reflects a non-hysteretic contribution, but in that case it corresponds to a saturation of the permittivity. 
The principal difference with the literature is the presence of a plateau before the increase to \qty{0}{\degree} which extends to larger electric fields here (\qty{20}{\kV\per\cm} instead of few \si{\kV\per\cm}).
This may be due to the different sample form (thin film here vs ceramics) [similar to Fig. S4 in \onlinecite{HashemizadehAPL2017}]. 
At \qty{325}{K}, the response is quite similar to the one at \qty{300}{\K}, even if the transition from \qty{-180}{\degree} to \qty{0}{\degree} is sharper.

For temperatures above $T_{\mathit{max}}$, i.e. above \SI{330}{K}, the phase-angle response exhibits a peculiar evolution: for low fields the value stays almost constant at \qty{-180}{\degree} but for higher fields instead of a transition by \qty{-90}{\degree} to \qty{0}{\degree}, the phase-angle decreases and exhibits a negative spike at \qty{-225}{\degree} followed by a positive spike at \qty{+45}{\degree}.
The crossing of $T_{\mathit{max}}$ and the associated change of regime is well visible on Fig.~\ref{subfig:delta TEMP OLEV map inc}.
This behavior has not been reported in the literature and this abrupt change in the phase-angle can be used to monitor the crossing of $T_{\mathit{max}}$.
A phase-angle of \qty{-270}{\degree} (or \qty{+90}{\degree}) indicates a pinching of the hysteresis loop\cite{MorozovJAP2008}.
For the pinching, a value close to \qty{-240}{\degree} is usually observed experimentally because of the competing contribution of the vibration from domain walls ($\delta_{3}=\qty{-180}{\degree}$)\cite{RiemerJAP2021}. 
Pinching of the hysteresis loop can have many origins: (i) fresh state for hard ferroelectrics such as Fe-doped \ce{PbZr_{0.58}Ti_{0.42}O3}\cite{MorozovJAP2008}, (ii) aged state and oxygen vacancies migration for Cu-doped \ce{BaTiO3}\cite{LiJECS2022}, Ce-doped \ce{Ba(Ti_{0.99}Mn_{0.01})O3}\cite{ZhaoCI2017} or (iii) electric field-induced transition from the paraelectric to a ferroelectric state for \ce{BaTiO3}\cite{MerzPR1953} or \ce{($1-x$)(Bi_{1/2}Na_{1/2})TiO3-$x$BaTiO3}\cite{SapperJAP2014}.
This later could play a role in our measurement since the change occurs near $T_{\mathit{max}}$ and the magnitude of the AC field stays low (and limits aging). 

The electric field $E_{t1}$ needed to induce a paraelectric to a ferroelectric state above $T_{\mathit{max}}$ should increase with the temperature \cite{SapperJAP2014,MerzPR1953}. 
One can note in our case, the electric field $E_{\mathit{AC}}^{\qty{-270}{\degree}}$, for which the phase angle of the third harmonic approaches \qty{-270}{\degree}, decreases from \qty{18}{\kV\per\cm} to \qty{12}{\kV\per\cm} when the temperature increases from \qty{335}{K} to \qty{380}{K}.
Thus, the field $E_{\mathit{AC}}^{\qty{-270}{\degree}}$ may not correspond to $E_{t1}$ but to the field for which (i) below, the non linearity is governed by reversible domain wall contribution ($\delta_{3}\simeq \qty{-180}{\degree}$) and (ii) above, the non-linearity is governed by saturation ($\delta_{3}\simeq \qty{0}{\degree}$). 
Around $E_{\mathit{AC}}^{\qty{-270}{\degree}}$, the deviation from the theoretical value, $\delta_{3}=\qty{-180}{\degree}$ for reversible and $\delta_{3}= \qty{0}{\degree}$ for saturation, is attributed to a pinching of the loop and the measured phase corresponds to the combination of the different contributions.\cite{RiemerJAP2021}
More explicitly, the negative spike at \qty{225}{\degree} results from a mix of reversible contributions and pinching, and the positive spike at \qty{+45}{\degree} corresponds to a mix between pinching and saturation.

At high temperatures (for \qty{375}{K} and especially for \qty{400}{K}) and low AC field ($E_{\mathit{AC}}<\qty{10}{\kV\per\cm}$), the non-linearity is so weak ($\Delta\varepsilon_{r}'<10$) that the phase angle of the third harmonic is difficult to measure resulting in a noisy behavior.

Directly following the measurements on heating, data have been acquired on cooling as well (Fig.~\ref{subfig:delta TEMP OLEV dec}). 
At \qty{375}{K}, the peculiar behavior ($\qty{-180}{\degree}\rightarrow \qty{-225}{\degree}\rightarrow \qty{+45}{\degree} \rightarrow \qty{0}{\degree}$) is less visible than on heating. 
At \qty{350}{K}, the peculiar behavior is not observed and instead the phase angle of the third harmonic evolution corresponds to a conventional relaxor behavior. 
At $T_{max}$ (\qty{330}{K}), the phase response corresponds to a relaxor but with a smoother transition from \qty{-180}{\degree} to \qty{0}{\degree} compared to data on heating. 
Below \qty{300}{K}, the phase evolution corresponds to a conventional ferroelectric with high field asymptotes from \qty{-30}{\degree} to \qty{-70}{\degree}.

\begin{figure*}
    \centering
    {\includegraphics[width=\textwidth]{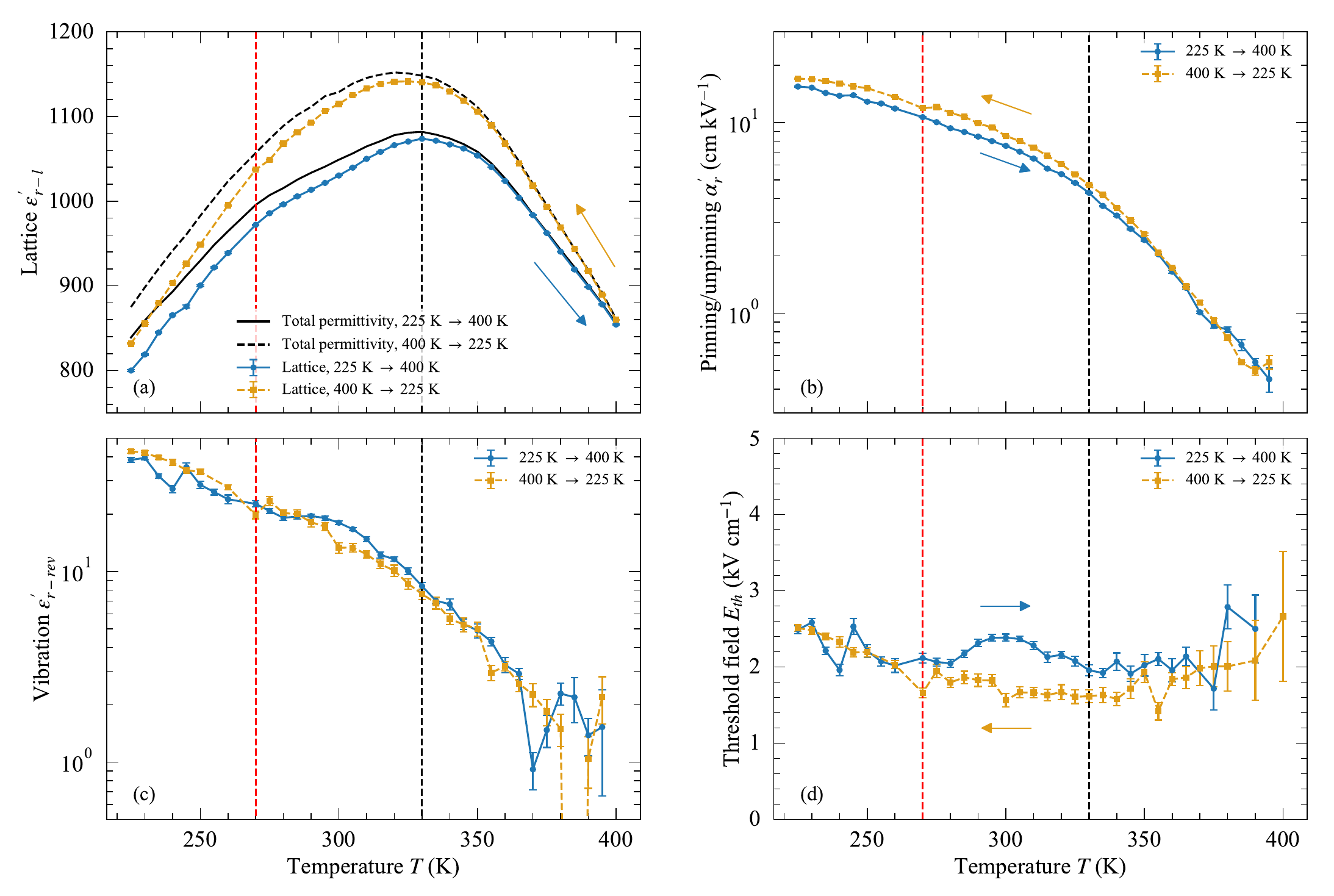}
    \subfloat{\label{subfig:realPerm Bulk TEMP}}%
    \subfloat{\label{subfig:Alph TEMP}}%
    \subfloat{\label{subfig:Reve TEMP}}%
    \subfloat{\label{subfig:Eth TEMP}}}%
    \caption{Relative permittivity for a measuring field of \qty{0.4}{\kV\per\cm}, lattice contribution to the permittivity (a), domain wall motion contributions to the permittivity (b,c) and threshold field (d) as a function of the temperature.}
    \label{fig:realPerm coeff TEMP}
\end{figure*}

These differences indicate that the domain structure (domains, domain walls, polar cluster boundaries, phase boundaries, etc) is different on heating and on cooling. 
Such difference can already be inferred from $P(E)$ hysteresis loops (see Fig.~S6b in supplementary material). 
At \qty{225}{K}, they exhibit a horizontal shift, characteristic of an internal field with a value around \qty{12}{\kV\per\cm}. 
It increases to around \qty{14}{\kV\per\cm} at \qty{300}{K} and then decreases with increasing temperatures. 
We postulate that this internal field indicates some correlations between PNRs, that would decrease with thermal fluctuations at higher temperatures. 
On cooling, the internal field is smaller (as low as \qty{6}{\kV\per\cm}) until \qty{325}{K}, implying that the correlations between PNRs are weaker. 
In the dielectric measurements we perform, at low temperatures, the application of the AC field may increase correlations between PNRs \cite{QiuJAP2019,TianAPL2023}. 
When the temperature increases, close to the phase transition, the correlation between the PNRs decreases due to thermal fluctuation (the internal bias field decreases) but when the AC field is applied, they can be correlated again, creating a pinching, i.e. a phase angle of the third harmonic approaching \qty{270}{\degree}. 
On cooling, the domain structure is different and most PNRs are uncorrelated since above the transition, the conventional relaxor behavior is observed and the AC field is not enough to induce pinching. 
On cooling further, PNRs stay less correlated and thus the global response of the material probed using AC field is closer to a conventional ferroelectric response.

The real part of the permittivity measurement as a function of the AC measuring field has been fitted using \eqref{hyperbolic} and Levenberg-Marquat algorithm and the resulting fits are shown as solid lines (Fig.~\ref{subfig:realPerm TEMP OLEV inc}).
Only the data below \qty{7}{\kV\per\cm} have been used for the fit since a deviation from the Rayleigh law occurs at higher fields.
The hyperbolic fits values are given in Fig.~\ref{fig:realPerm coeff TEMP} as a function of the temperature, with associated confidence intervals.
The lattice contribution shows a similar evolution with temperature as the total permittivity, consistent with the fact that, at low fields, it represents the main contribution to the permittivity.
Its maximum occurs for \qty{334}{K}, very close to the $T_{\mathit{max}}$ found at low fields (\qty{330}{\K}).
The slight difference below $T_{\mathit{max}}$ correspond to the reversible domain wall motion contribution.

Fig.~\ref{subfig:Alph TEMP},\subref*{subfig:Reve TEMP} show the domain wall motion contributions to the permittivity. 
There a no measurements in the literature on thin films of BCTZ. 
Still it is useful to compare with room temperature measurements on ceramics where $\alpha_{r}$ is around \qty{300}{\cm\per\kV} to \qty{3000}{\cm\per\kV}\cite{LiuAMI2017,AbebeJAP2017,ZhengAPL2023}, much larger than the values we find here. 
This indicates that domain walls are less mobile in thin films which is attributed to clamping of the film \cite{KeechJAP2014}. 

Below $T_{\mathit{max}}$, both domain wall motion contributions decrease when the temperature increases, without any change at 270 K where there was an inflection point in the real part of the permittivity. 
This differs from what has been found for films of \ce{(Pb,Sr)TiO3}\cite{BaiCI2017} or \ce{0.5PbYb_{1/2}Nb_{1/2}-0.5PbTiO3}\cite{BassiriGharbJE2007} for which the domain wall motion contribution increases when the temperature approaches $T_{\mathit{max}}$.
Above $T_{\mathit{max}}$, the domain wall motion coefficients still decrease but are not null, corresponding to a residual ferroelectricity\cite{GartenJAP2014,GartenJACS2016}, which persists here \qty{70}{\K} above $T_{\mathit{max}}$.
The decrease is slightly more pronounced than before: at \qty{330}{\K} $\alpha_{r}'$ is 3 times lower than at \qty{260}{\K} whereas at \qty{400}{\K}, $\alpha_{r}'$ is 9 times lower than a \qty{330}{\K}.
The change of decay rate is also visible on the reversible contribution.
According to Boser\cite{boserjap1987}, the reversible domain wall contribution is proportional to the domain wall density and the distance traveled by the domain wall, this later depending on the potential energy profile of the domain wall\cite{BassiriGharbJE2007}. 
A stronger decrease of the reversible contribution could thus indicate faster changes in the density of domain walls, which would be consistent with the fact that above $T_{\mathit{max}}$ the number of PNRs and their sizes both decrease.\cite{MaAPL2013,YeAPL2015} 
In the studied sample, the reversible and irreversible contributions do not exhibit a peak, contrary to what has been seen for \ce{NaNbO3} \cite{CaiPRB2016}, \ce{Pb_{0.92}La_{0.08}Zr_{0.52}Ti_{0.48}O3} \cite{MaAPL2013}, Nb-doped \ce{PbZrO3} \cite{YeAPL2015} or \ce{PbMn_{1/3}Nb_{2/3}O3}\cite{ShettyAFM2019}.

One can note that the reversible domain wall contribution fluctuates above \qty{350}{K} because the non-linearity becomes very weak above $T_{max}$ and makes the coefficient extraction using \eqref{hyperbolic} difficult. 
The confidence intervals in absolute values are similar for the lattice contribution and the reversible contribution (mean value of 0.6). This effect is accentuated by the semi-log scale

Using reversible and irreversible domain wall motion contributions, it is possible to determine the threshold field $E_{\mathit{th}} = \varepsilon_{\mathit{r-rev}}'/\alpha_{r}'$ which represents the degree of pinning of the domain walls\cite{borderonapl2011,BorderonSR2017,GharbJAP2005}.
The threshold field remains almost constant when the temperature changes, indicating a constant degree of pinning.
This suggests that the decrease of the non-linear behavior, i.e. the irreversible domain wall motion contribution, with increasing temperature results from a decrease of the domain wall density and not a change in energy profile. 
The fluctuation at temperatures above \qty{350}{K} results from fluctuations of the reversible domain wall motions contribution. 

When repeating the measurement on cooling, the lattice contribution is slightly higher (similar to what is found for the total permittivity). 
The irreversible domain wall contribution is also slightly higher ($\qty{15}{\%}$) indicating a higher non-linear behavior. 
Since the reversible domain wall contribution is similar on cooling and on heating, the higher irreversible contribution is due to higher domain wall mobility, which is confirmed by a small value of the threshold field on cooling. 

In summary, in this article, we studied the sub-coercive field non-linearities as a function of the temperature of a BCTZ 50/50 thin film.
The relaxor behavior of the film is confirmed by the shift of the maximum permittivity temperature $T_{\mathit{max}}$ with frequency and the slim shape of the $P(E)$ loops.
We have shown that $T_{\mathit{max}}$ also depends on the AC measuring field because of the irreversible domain wall motion contribution.
The domain wall contributions is found to regularly decrease with temperature and still contribute to the permittivity above $T_{\mathit{max}}$.
In addition, phase angle of the third harmonic measurements show that the thin film behaves like a conventional ferroelectric below \qty{275}{K} and as a relaxor from \qty{275}{\K} to $T_{\mathit{max}}=\qty{330}{\K}$.
Above $T_{\mathit{max}}$, the thin film exhibits a peculiar phase angle of the third harmonic response which consists of $\qty{-180}{\degree}\rightarrow \qty{-225}{\degree}\rightarrow \qty{+45}{\degree} \rightarrow \qty{0}{\degree}$ instead of the $\qty{-180}{\degree}\rightarrow \qty{-90}{\degree} \rightarrow \qty{0}{\degree}$ found for relaxor.
This peculiar behavior is observed only on heating, and is tentatively attributed to changes in the correlations between polar nanoregions.

\section*{Supplementary Material}
The thin film growth procedure, the associated phase/composition characterizations, polarization versus electric field loop and modified Curie-Weiss analysis are given in supplementary material.

\section*{Data availability}
The data that support the findings of this study are available from the corresponding author upon reasonable request.

\section*{Acknowledgments}
This work has been performed with the means of the CERTeM (microelectronics technological research and development center) of French region Centre Val de Loire.
This work was funded through the project MAPS in the program ARD+ CERTeM 5.0 by the Région Centre Val de Loire co-funded by the European Union (ERC, DYNAMHEAT, N°101077402). 
Views and opinions expressed are however those of the authors only and do not necessarily reflect those of the European Union or the European Research Council. 
Neither the European Union nor the granting authority can be held responsible for them.

The authors would like to thank Brahim Dkhil for fruitful discussions on the relaxor behavior.

\section*{Conflict of Interest}
The authors declare no competing financial interest.

\bibliographystyle{aipnum4-2}
\bibliography{biblio_ferro.bib}
\end{document}